\documentstyle[12pt,amsfonts,amssymb]{article}
\def\emph#1{{\em#1}}
\def\textbf#1{{\bf#1}}
\def\textit#1{{\it#1}}
\def\textsf#1{{\sf#1}}
\def\texttt#1{{\tt#1}}

\scrollmode %% don't stop for errors

\advance\hoffset by -1.1truecm
\advance\voffset by -1.0truecm

\renewcommand{\thefootnote}{\fnsymbol{footnote}}

\newcommand{\be}{\begin{equation}}
\newcommand{\ee}{\end{equation}}
\newcommand{\bea}{\begin{eqnarray}}
\newcommand{\eea}{\end{eqnarray}}
\newcommand{\bean}{\begin{eqnarray*}}
\newcommand{\eean}{\end{eqnarray*}}
\newcommand{\bb}{\begin{eqnarray}}
\newcommand{\eee}{\nonumber\end{eqnarray}}

\renewcommand{\AA}{{\mathbb A}} %% NC gauge potential
\newcommand{\C}{{\mathbb C}} %% complex numbers
\newcommand{\F}{{\mathbb F}} %% NC gauge field
\newcommand{\HH}{{\mathbb H}} %% quaternions
\newcommand{\R}{{\mathbb R}} %% real numbers
\newcommand{\opname}[1]{\mathop{\rm #1}\nolimits} %%_\det,\ker,...

\newcommand{\Tr}{\opname{Tr}} %% trace
\newcommand{\tr}{\opname{tr}} %% trace of matrix
\newcommand{\A}{{\cal A}} %% an algebra
\newcommand{\stroke}{\mathbin{\vert}} %% (for `\braket' and such)
\newcommand{\braket}[2]{\langle#1\stroke#2\rangle} %% Dirac bracket
\newcommand{\delslash}{{\partial\mkern-9mu/}} %% usual Dirac operator
\newcommand{\eps}{\epsilon} %% abbreviation for \epsilon
\newcommand{\ga}{\gamma} %% abbreviation for \gamma
\renewcommand{\H}{{\cal H}} %% Hilbert space
\newcommand{\la}{\lambda} %% abbreviation for \lambda
\newcommand{\ox}{\otimes} %% tensor product
\newcommand{\pa}{\partial} %% abbreviation for \partial
\newcommand{\pp}[1]{\pmatrix{#1}} %% abbreviation for \pmatrix
\newcommand{\qq}{\quad} %% abbreviation for \quad
\newcommand{\tfrac}[2]{{\textstyle\frac{#1}{#2}}} %% small fraction
\newcommand{\thalf}{{\textstyle\frac{1}{2}}} %% small fraction 1/2
\newcommand{\x}{\times} %% cartesian or cross product
\newcommand{\7}{\dagger} %% abbreviation for + symbol

\newbox\ncintdbox \newbox\ncinttbox
\setbox0=\hbox{$-$}
\setbox2=\hbox{$\displaystyle\int$}
\setbox\ncintdbox=\hbox{\rlap{\hbox
to \wd2{\hskip-.125em\box2\relax\hfil}}\box0\kern.1em}
\setbox0=\hbox{$\vcenter{\hrule width 4pt}$}
\setbox2=\hbox{$\textstyle\int$}
\setbox\ncinttbox=\hbox{\rlap{\hbox
to \wd2{\hskip-.175em\box2\relax\hfil}}\box0\kern.1em}

\newcommand{\ncint}{\mathop{\mathchoice{\copy\ncintdbox}%
{\copy\ncinttbox}{\copy\ncinttbox}{\copy\ncinttbox}}\nolimits}

\begin{document}

\begin{titlepage}

\begin{center}

\textbf{Centre de Physique Th\'eorique, CNRS -- Luminy, Case 907}

\textbf{F--13288 Marseille -- Cedex 9}

\end{center}

\vspace{2cm}

\begin{center}

\setcounter{footnote}{0}
\renewcommand{\thefootnote}{\arabic{footnote}}

{\large\bf THE STANDARD MODEL IN NONCOMMUTATIVE\\[6pt]
GEOMETRY AND FERMION DOUBLING}

\vspace{1cm}

\textbf{J. M. GRACIA-BOND\'IA,%
\footnote{Departamento de Matem\'aticas, Universidad de Costa Rica,
San Pedro 2060, Costa Rica and
Departamento de F\'{\i}sica Te\'orica, Universidad de Zaragoza,
50009, Zaragoza, Spain.}\\[3pt]
B. IOCHUM and T. SCH\"UCKER%
\footnote{Universit\'e de Provence and Centre de Physique Th\'eorique,
CNRS--Luminy, Case 907,
13288 Marseille, France.
E-mail: \texttt{iochum@cpt.univ-mrs.fr},
\texttt{schucker@cpt.univ-mrs.fr}}}

\vspace{2.3 cm}

\textbf{Abstract}
\end{center}
The link between chirality in the fermion sector and
(anti-)self-duality in the boson sector is reexamined in the light of
Connes' noncommutative geometry approach to the Standard Model. We
find it to impose that the noncommutative Yang--Mills action be
symmetrized in an analogous way to the Dirac--Yukawa operator itself.

\vspace{2 cm}

\noindent July 1997

\noindent CPT--97/P.3503
\smallskip

\noindent Other preprint Nos.: DFTUZ/97/08, UCR--FM--11--97,
hep-th\slash 9709145.

\bigskip

\noindent anonymous ftp or gopher: \texttt{cpt.univ-mrs.fr}

\end{titlepage}

\setcounter{footnote}{0}
\renewcommand{\thefootnote}{\arabic{footnote}}

\subsection*{1. Introduction}

The improved Connes--Lott~\cite{RealNCG} and
Chamseddine--Connes~\cite{ChamC-PRL, CIKS, Zappafrank} models of
noncommutative geometry (NCG) have already yielded action functionals,
respectively for elementary particles (tying together the gauge bosons
and the Higgs sector) and for elementary particles coupled to gravity.
In both kinds of models, the generalization of the Yang--Mills (and
gravitational, as the case may be) action is obtained from an
operator-theoretic data set: $(\A, \H, D, J,\chi)$, where $\A$ is a
suitable noncommutative algebra, replacing the algebra of functions on
ordinary spacetime, acting on the linear space $\H$, related to the
fermionic content of the theory, $D$ is a generalized Dirac operator,
involving both spacetime and internal degrees of freedom, $J$ is a
charge conjugation operator (making the triple $(\A, \H, D)$ into a
{\it real\/} spectral triple or $K$-cycle) and $\chi$ is the
chirality.

This reinterpretation of the Standard Model (SM) of fundamental
interactions as an unveiling of the (noncommutative) geometric
structure of the spacetime has many intriguing aspects. The advantages
afforded by the tools of (this variant of) NCG over alternative
descriptions have been discussed elsewhere~\cite{Cordelia}.
Nevertheless, Lizzi~{\it et al\/}~\cite{DoubleTrouble} ---see
also~\cite{Mirrorworld}--- recently pointed out that the Hilbert
spaces employed till now in the NCG approach overcount the fermion
degrees of freedom. They stated that a ``physical'' Hilbert space
could be obtained by a sort of chiral projection of the Hilbert space
pertaining to the product $K$-cycle of~\cite{RealNCG, ChamC-PRL}; but
then, according to them, when retaining only the contribution of the
physical states, the NCG procedure to recover the bosonic part of the
action fails.

In~\cite{DoubleTrouble}, as in most NCG papers, calculations are made
in the Euclidean framework, whereas conclusions are drawn on the SM
Lagrangian, in a different spacetime signature. Therefore, in
discussing claims such as the ones made by Lizzi~{\it et al\/}, it
seems a useful start, to clarify the procedure employed in NCG for
mimicking the Lorentzian framework: on that we only know of a brief
remark in all the pertinent literature~\cite{CLCargese}. We devote the
next section to an account of that procedure, which we feel takes out
the force of {\it some of\/} the arguments in~\cite{DoubleTrouble};
also we briefly discuss the variants of a relative sign apparently
introduced thereby in the fermion sector of the Lagrangian.

On the other hand, the reasoning by Lizzi~{\it et al\/} uncovers an
interesting link between chirality in the fermion sector and
(anti-)self-duality in the boson sector; to wit, projection over
states of definite chirality in the source happens to lead to
projection onto eigenstates of the signature operator in the gauge
fields. It is a deep result of noncommutative
geometry~\cite{ConnesBreak} that all of the geometrical structure of a
Riemannian spin manifold can be recovered from a set of axioms for
commutative real spectral triples. A similar theorem for Lorentzian
spin manifolds (and thus a {\it prima facie} Lorentzian formulation of
the Connes--Lott and Chamseddine--Connes models of noncommutative
geometry) as yet does not exist, and, to our minds, it may demand
abandoning the (tensor) {\it product spectral triples} till now
favoured in the literature. At any rate, however, the axioms of
algebraic character (in particular, the crucial condition of
commutativity between $[D, \A]$ and $J\A J^\7$) among that set, can be
straightforwardly reformulated in the Lorentzian context; and so can
the algebraic constructs leading to the Connes--Lott type of models,
provided one is willing to trade Krein spaces for Hilbert spaces and
work with pseudoscalar products~\cite{Cordelia}. Now, the link between
chirality and self-duality seems robust enough to be essentially
independent of the signature ---and there, the arguments
of~\cite{DoubleTrouble} cannot be so easily dismissed.

In effect, we show in section 3 that the troubles pointed out
in~\cite{DoubleTrouble} are a blessing in disguise. The aforementioned
link motivates us to plant signposts in the way of the future
Lorentzian formulation of the theory, and also helps to decide a
relatively subtle question in real spectral triple theory. In the
third section of the paper, we show that the bosonic part of the SM
Lagrangian can be reconstructed in the Connes--Lott way, even when the
chiral projection of~\cite{DoubleTrouble} is made, provided that the
{\it symmetrized\/} inner product on noncommutative differential forms
(seen as operators on the pertinent linear space) is adopted; i.e.,
provided one takes into account not only the given representation of
$\A$, but also its conjugate representation by $J$ ---which Lizzi
{\it et al\/} omitted to do.

\subsection*{2. Minkowskian versus Euclidean}

We start by recalling the noncommutative spectral data for the SM,
following~\cite{Cordelia}. Since the T\v{r}e\v{s}\v{t} conference of
May~95, it has become standard in the NCG literature to identify the
arena for the SM as the algebra
$$
\A_t := C^\infty(M,\R) \ox \A_f,
$$
where $M$ denotes the ordinary spacetime and $\A_f$ the finite
dimensional real Eigenschaften algebra:
$$
\A_f := \HH \,\oplus\, \C \,\oplus\, M_3(\C),
$$
involving in particular the algebra $\HH$ of quaternions.

In the product spectral triple formulation, the Hilbert space $\H_t$
is the tensor product of a generic {\it bispinor\/} space $L^2(S_M)$
by a 90-dimensional complex space $\H_f$ with a basis labelled by all
chiral particles and antiparticles in the SM; hence the apparent
redundancy of degrees of freedom. An element of $\H$ is written as a
multispinor field
$\pmatrix{\Psi_L \cr \Psi_R \cr \Psi_L^c \cr \Psi_R^c}$. We have
indicated that $\A$ acts on $\H$. The (real) representation $\pi$ of
$\A$ in the lepton sector is given by
$$
\pi(q,\la,m) \pmatrix{\Psi_L \cr \Psi_R \cr \Psi_L^c \cr \Psi_R^c}
= \pmatrix{q \Psi_L \cr \la \Psi_R \cr
\bar\la \Psi_L^c \cr \bar\la \Psi_R^c \cr}.
$$
In the quark sector:
$$
\pi(q,\la,m) \pmatrix{\Psi_L \cr \Psi_R \cr \Psi_L^c \cr \Psi_R^c}
= \pmatrix{q \Psi_L \cr \pi^q(\la) \Psi_R \cr
m \Psi_L^c \cr m \Psi_R^c \cr},
$$
where $m$ acts on the internal color space of each quark field and
$\pi^q$ is the real representation of $C^\infty(M,\C)$ acting as
$\bar\la$ on the $u$-labelled quark fields and as $\la$ on the
$d$-labelled quark fields. Moreover, if we write
$D_t = \delslash \ox 1 + \gamma_5 \ox D_f$, $J_t = C \ox J_f$, where
$C$ denotes charge conjugation, and $\chi_t = \gamma_5 \ox \chi_f$
where $\chi_f$ is the natural grading on $\H_f$, then
$(\A_t, \H_t, D_t, J_t, \chi_t)$ is a real spectral triple.

We wish to remark~\cite{CammaCoq} that the quaternions are already
present in the usual formalism of the Standard Model. This is
tantamount to the $SU(2)$ gauge invariance of the mass terms ---see,
for instance~\cite{DonKerson}. The fact is not always recognized or
exploited.

The ``covariant'' Dirac operator, restricted to the particle space, is
schematically written:
$$
D_p = \pmatrix{\ga^\mu (i\partial_\mu + L_\mu) & \gamma_5 M \cr
\gamma_5 M^\7 & \ga^\mu(i\partial_\mu + R_\mu)},
$$
where, for instance in the quark sector,
$$
M = \pmatrix{\bar\phi^0 & \phi^+ \cr -\phi^- & \phi^0}
\pmatrix{m_u & \cr & m_d},
$$
with the $3 \x 3$ (where 3 is the number of fermionic generations and
we are forgetting here about color) matrices $m_u, m_d$ encoding all
the dimensionless Yukawa couplings and Kobayashi--Maskawa mixing
parameters, and the mass matrix $M_0$ could be exhibited by shifting
$\phi^0$, $\bar\phi^0$ by a constant with the dimensions of the scalar
field (i.e., mass), corresponding to its vacuum expectation value. The
$L_\mu$ and $R_\mu$ collectively denote the gauge fields coupled to
each chiral sector, which {\it together with the Higgs field}, in NCG
are determined by the interplay between the Dirac operator and the
world algebra action: the full Dirac--Yukawa operator on the space of
particles and antiparticles is $D = D_p + JD_pJ^\7$ and the coupling
of fermions to all boson fields is described by $\AA + J\AA J^\7$,
where $\AA$ denotes the noncommutative gauge potential (a Hermitian,
noncommutative 1-form) associated to the triple $(\A, \H, D)$. In
order that our chosen data set $(\A, \H, D, J)$ be accepted as a
certified real spectral triple, one needs to check that both $\A$ and
$[D,\A]$ commute with the ``conjugate action'' $J\A J^\7$ and
reciprocally~\cite{ConnesBreak}.

Note that the choice of $D_t$ is not unique; for instance,
$D_t = U(\delslash \ox \chi_f + 1 \ox D_f)U^\7$, where U is the unitary
$U = P_+ \ox 1 + P_- \ox \chi_f$ with
$P_{\pm} = \thalf(1 \pm \gamma_5)$ is another possible choice. This
shows that the relative sign between the mass terms introduced by the
presence of $\gamma_5$
in $D_t$ can be replaced by another one between kinetic terms when the
internal chirality
$\chi_f$ is prefered and has nothing to do with a choice of signature.

\smallskip

Let us briefly review the old story of chiral fermion doubling in the
Euclidean. In the Minkowskian, for one species of fermions, nature has
elected the Dirac Lagrangian,
\bb
\overline{\Psi_L}\gamma^\mu \,i\pa_\mu\Psi_L +
\overline{\Psi_R}\gamma^\mu \,i\pa_\mu\Psi_R,
\eee
where $\Psi_L$ and $\Psi_R$ are two component chiral spinors,
\bb
\Psi_L := \,\frac{1-\gamma_5}{2}\, \Psi,  \qq
\Psi_R := \,\frac{1+\gamma_5}{2}\, \Psi,
\eee
and the pseudoscalar product $\overline\Psi\Psi$ is defined by
$\overline \Psi := \Psi^\7\gamma^0$. This pseudoscalar product is
motivated physically by probability conservation, mathematically by
invariance under the Clifford algebra. For concreteness we specify our
Dirac matrices:
\bb
\gamma^0 = \pp{0 & 1 \cr 1 & 0},  \qq
\gamma^k = \pp{0 & \sigma_k \cr -\sigma_k & 0},  \qq  k = 1,2,3,
\eee
with the Pauli matrices
\bb
\sigma_1 = \pp{0 & 1 \cr 1 & 0},  \qq
\sigma_2 = \pp{0 &-i \cr i & 0},  \qq
\sigma_3 = \pp{1 & 0 \cr 0 &-1}.
\eee
The Dirac matrices satisfy the anticommutation relations,
\bb
\gamma^\mu\gamma^\nu + \gamma^\nu\gamma^\mu = 2\eta^{\mu\nu} 1_4, \qq
\eta^{\mu\nu} = \opname{diag}(1,-1,-1,-1).
\eee
Chirality is then the unitary operator of unit square
\bb
\gamma_5 = \frac{i}{4!} \epsilon_{\mu\nu\rho\sigma}
\gamma^\mu\gamma^\nu\gamma^\rho\gamma^\sigma
= i\gamma^0\gamma^1\gamma^2\gamma^3 = \pp{-1 & 0 \cr 0 & 1},  \qq
\epsilon_{0123}=1.
\eee
It anticommutes with all four $\gamma^\mu$. Charge conjugation is the
antiunitary operator of unit square
\bb
J\Psi = \Psi^c= i\gamma^2 \Psi^*.
\eee
It anticommutes with all four $\gamma^\mu$.

\smallskip

The invariant scalar product in the Euclidean $\Psi^\7\Psi$ has no
$\gamma^0$ and the corresponding Euclidean Lagrangian,
\bb
(\Psi_L)^\7 \gamma^\mu \,i\pa_\mu \Psi_L +
(\Psi_R)^\7 \gamma^\mu \,i\pa_\mu \Psi_R,
\eee
vanishes identically. The way out is to assume that $\Psi_L$ and
$\Psi_R$ are two independent 4-spinors. This doubling of degrees of
freedom is usually taken as an artifact of the Euclidean formulation
and at the end of day one ``Wick rotates'' back to the Minkowskian and
then imposes the chirality conditions,
\bb
\frac{1-\gamma_5}{2}\,\Psi_L = \Psi_L,  \qq
\frac{1+\gamma_5}{2}\,\Psi_R = \Psi_R.
\eee
In~\cite{Mirrorworld}, Lizzi {\it et al} propose to grant physical
significance to the additional ``mirror'' fermions.

\subsection*{3. The physical inner product}

The noncommutative integral allows one to define scalar or
pseudoscalar products on the algebra of noncommutative differential
forms. The prototype would be
$$
\ncint S^\7 T := \Tr_{\rm Dix}\,(S^\7 T D^{-4}),
$$
where $\Tr_{\rm Dix}$ denotes the Dixmier trace. It is known that,
under suitably regularity conditions, $\ncint$ defines a trace on that
algebra~\cite{Sigfrid, Ciprianietal}; it resolves in our cases into a
combination of finite traces and ordinary integrals~\cite{Cordelia};
the final integrand is to be interpreted as the Lagrangian.

Let $\F = \F^\7$ denote the noncommutative gauge field associated to
$\AA$ (for our algebra $\A_t$) and let $P_+$ be the projection over
the ``physical subspace'', in the language of~\cite{DoubleTrouble}.
That is, $P_+$ is equivalent to $\thalf(1 + \gamma_5)$ on the
subspaces of $H_f$ labelled ``right'' and to $\thalf(1 - \gamma_5)$ on
the subspaces of $H_f$ labelled ``left''. Then the main contention of
that paper is that $\ncint P_+ \F^2$ yields in that integrand only the
anti-self-dual component of the kinetic term of the $SU(2)$ field:
$$
C_F\,\tr(F^{\mu\nu}F_{\mu\nu} - F^{\mu\nu}{\ast F_{\mu\nu}}),
$$
where, as usual,
$\ast F_{\mu\nu} = \thalf\eps^{\mu\nu\sigma\rho} F_{\sigma\rho}$.
Analogously, for the $U(1)$ field they get the self-dual component:
$$
C_B(B^{\mu\nu}B_{\mu\nu} + B^{\mu\nu}{\ast B_{\mu\nu}}).
$$
For the color field, the usual kinetic term is obtained. Note that the
corresponding classical Yang--Mills equations of motion are not
modified by the addition of the topological term; but the result is
presumably not acceptable on quantum grounds.

With some care, it is feasible to recast the Connes-Lott procedure in
the Lorentzian mould. We give now a parallel account of the procedure
in~\cite{DoubleTrouble}.

\smallskip

Let $F = \thalf F_{\mu\nu} \gamma^\mu\gamma^\nu$ be a Hermitian
2-form. The starting point is the scalar product in component form:
\bb
\ncint F^2
= \frac{1}{128\pi^2} \int_M F_{\mu\nu} F_{\rho\sigma}
\tr[\gamma^\mu\gamma^\nu\gamma^\rho\gamma^\sigma]
= \frac{1}{16\pi^2} \int_M F^{\mu\nu} F_{\mu\nu}
= \frac{1}{8\pi^2} \int_M F\,{\ast F}.
\eee
Motivated from the chiral projection $P_\pm$ in the fermionic sector,
Lizzi {\it et al} also project in the bosonic sector. With
\bb
\ncint\gamma_5 F^2
= \frac{1}{128\pi^2} \int_M \! F_{\mu\nu} F_{\rho\sigma}
\tr[\gamma_5\gamma^\mu\gamma^\nu\gamma^\rho\gamma^\sigma]
= \frac{i}{32\pi^2} \int_M \! F_{\mu\nu} F_{\rho\sigma}
\epsilon^{\mu\nu\rho\sigma} = \frac{i}{8\pi^2} \int_M F F,
\eee
we get the Yang--Mills action plus the well known topological term
\bb
\ncint (P_\pm F)^2 = \ncint P_\pm F^2 = \frac{1}{32\pi^2}
\int_M (F^{\mu\nu}F_{\mu\nu} \pm i F^{\mu\nu}{\ast F_{\mu\nu}})
= \frac{1}{16\pi^2} \int_M(F{\ast F} \pm iFF).
\eee

\smallskip

This result is neither unsatisfactory nor wholly unexpected. The
situation is reminiscent of the fact that the chiral projection
applied to the gravitational action in Schwinger's formulation
\cite{schw} produces the gravitational action in Ashtekar's
formulation \cite{htk}. This is common folklore, but we sketch the
proof below, because it is not often recognized or exploited by
noncommutative geometers. Just write the (Minkowskian) Dirac operator
in components,
\bb
D = \gamma^a e^\mu_a (\pa_\mu + \tfrac{1}{4} \omega_\mu), \qq
\omega_\mu := \tfrac{1}{2} \omega_{ab\mu} [\gamma^a,\gamma^b],
\eee
where $\omega_{ab\mu}$ are the components of the Levi-Civita
connection with respect to the orthonormal frame $e^\mu_a\pa_\mu$.
After a $1 + 3$ split, $a = (0,j) = (0,1,2,3)$, we have
\bb
\omega_\mu = -i{\epsilon_{0ij}}^k
\pp{^{(+)}\chi_{ij\mu}\sigma_k & 0 \cr 0 & ^{(-)}\chi_{ij\mu}\sigma_k},
\eee
where
\bb
^{(\pm)}\chi_{ij\mu}
= \omega_{ij\mu}\,\mp\, i{\epsilon_{0ij}}^\ell \,\omega_{0\ell\mu}
\eee
are Ashtekar's variables.

The foregoing is related to the fact that the Lorentz group and the
special orthogonal group in three complex dimensions are locally
isomorphic, that again seems also widely ignored. One would expect, as
suggested at the end of reference~\cite{ChinosQuirales} in the context
of effective actions, that the Chamseddine--Connes action computed
from a truly chiral theory would yield an (anti-)self-dual formulation
of gravity.

Now, it stands to reason that, in such a context, the appropriate $J$
is the charge conjugation operator for the fermion multiplet of the
SM, with suitable phase factors~\cite{Scheckbook}. The crucial
observation, from the viewpoint of this paper, is that in the
Minkowskian $J$ interchanges particles of {\it opposite\/} chirality,
that is, $(\Psi_L)^c = (\Psi^c)_R$ is a right-handed spinor and so on.

We assert that, in the same way as the covariant Dirac operator
incorporates the symmetrized noncommutative gauge potential
$\AA + J\AA J^\7$, the correct inner product formula defining the
generalized Yang--Mills functional incorporates the symmetrized
noncommutative gauge field:
$$
I_{YM}(\F) := \ncint (\F + J\F J^\7)^2.
$$
If $\A$ is commutative, then $J\AA J^\7 = -\AA^\7$. Thus, for a
Hermitian $\AA$, we have $\AA + J\AA J^\7 = 0$ in that case. However,
for the SM, this scalar product is nondegenerate. The symmetrization
process is justified by the spectral principle of~\cite{ChamC-PRL}:
$D_t + \AA$ is forbidden since for any unitary $u \in \A_t$, the inner
automorphism $\alpha_u\colon a \in \A_t \mapsto uau^\7 \in \A_t$
induces a unitary operator $U = \pi_t(u)J\pi_t(u)J^\7$ satisfying
$U \pi_t(a) U^\7 = \pi_t(\alpha_u(a))$ and
$UD_tU^\7 = D_t + \AA + J\AA J^\7$ with $\AA = \pi_t(u)[D_t, \pi_t(u)^\7]$.
So $D_t$
and $D_t + \AA + J\AA J^\7$ have the same spectrum.

Also, were one to proceed otherwise, one would be treating the
contributions of the direct and the conjugate representation (and then
of particles and antiparticles) on a different footing, in spite of
the fact that their r\^oles are interchangeable.

We show that, due to the Minkowskian relation
$\gamma_5 J = - J\gamma_5$, the pertinent contributions of the
$J \F^2 J^\7$ term are of the form
$C_F\,\tr(F^{\mu\nu}F_{\mu\nu} + F^{\mu\nu}{\ast F_{\mu\nu}})
+ C_B (B^{\mu\nu}B_{\mu\nu} - B^{\mu\nu}{\ast B_{\mu\nu}})$, matching
the previous ones in such a way that the usual SM action is recovered.
Indeed,
\bb
\ncint (P_+ F + J P_+F J^\7)^2
= \ncint (P_+ F)^2 + \ncint (P_- F)^2 + 0 = \ncint F^2.
\eee

\smallskip

To summarize: Connes' spectral principle makes the symmetrized scalar
product mandatory. Lizzi~{\it et al\/} uncovered an independent
argument in favour of this symmetrized product:
$$
\braket{S^\7}{T} := \Re e \ncint (S + JSJ^\7)^\7 (T + JTJ^\7)
$$
on the {\it real\/} differential algebra of the real spectral triples.

Besides the SM, in~\cite{DoubleTrouble} the noncommutative bosonic
action is computed for another model, in which there are
two different $SU(2)$ gauge fields, each one coupling to each chiral
sector. They as well obtain, in their Euclidean framework, both in the
Connes--Lott and in the Chamseddine--Connes temperaments, that the
projection over states of definite chirality on the fermionic source
leads in NCG to a similar projection in the gauge fields. The results
of Lizzi~{\it et al\/} for that toy model of course stand, but let us
observe that it is not endowed with a charge conjugation operator
---and in that sense perhaps does not describe a true noncommutative
geometry.

\smallskip

Our argument hinges on the fact that antiparticles have the opposite
chirality to their corresponding particles. It stands for the
present in a sort of no man's land, as the just-mentioned property is
of course murky in the Euclidean formalism, the only one in which
rigorous NCG calculations to date have been performed: to give a fully
Lorentzian version of the NCG action principle is a formidable task,
that we leave for another day.

The NCG models of the general type considered in this paper give
constrained versions of the SM, i.e., with some relations among the
parameters of the boson sector, allowing, namely, to make educated
guesses on the Higgs particle mass. Those coefficients vary when
computed with the symmetrized inner product. The computation is not
entirely straightforward, chiefly because of the well known
nonlinearity that arises (in the Connes--Lott procedure) when
combining the contribution of the quark and the lepton sectors. That
issue has no bearing on the main point of this article and has been
reported separately~\cite{CIKS2}.

\vspace{2cm}

\subsubsection*{Acknowledgments}

\medskip

JMG-B acknowledges support from the Universidad de Costa Rica and thanks
the Departamento de F\'{\i}sica Te\'orica of the Universidad de Zaragoza
and the Centre de Physique Th\'eorique (CNRS--Luminy) for their warm
hospitality.  He also thanks L.~J.~Boya for reminding him of the fact
that the Lorentz group is essentially the orthogonal group in three
complex dimensions, F. Lizzi for useful suggestions and J.~C.~V\'arilly
for discussions on the symmetries of the SM. We three are grateful to
A.~Connes, W.~Kalau, D.~Kastler and T.~Krajewski for illuminating exchanges of
views concerning the issues touched upon in this paper.

\end{document}